# Panchromatic Mining for Quasars: An NVO Keystone Science Application

Robert J. Brunner, Palomar Observatory, California Institute of Technology

**ABSTRACT**

A data Tsunami is overwhelming Astronomy. This wave is affecting all aspects of our field, revolutionizing not just the type of scientific questions being asked, but the very nature of how the answers are uncovered. In this invited proceeding, we will address a particular scientific application — Panchromatic Mining for Quasars — of the forthcoming virtual observatories, which have arisen in an effort to control the effects of the data Tsunami. This project, in addition to serving as an important scientific driver for virtual observatory technologies, is designed to a) characterize the multi-wavelength nature of known active galaxies and quasars, especially in relation to their local environment, in order to b) quantify the clustering of these known systems in the multidimensional parameter space formed by their observables, so that new, and potentially unknown types of systems can be optimally targeted.

## 1. Introduction

In this invited proceedings, we introduce the Skyserver project, which is federating all publicly available, large astronomical catalogs. One of the problems that we face as a consequence of the digital revolution is the amazing wealth of data that we now posses. However, we now know that data possession and data utilization are often extremely uncorrelated, and that we need to rethink how we store, handle and analyze the large quantities of data that are becoming publicly available. In general, solutions to this problem fall under the umbrella term of virtual observatories.

In this contribution, we first discuss the keystone virtual observatory science driver of panchromatic quasar mining, including a list of specific astrophysical questions that we wish to tackle. Following that, we delve into the challenges that must be overcome due to the size and complexity of the data involved. Next, we present a detailed discussion of the relevant technical aspects of the project. Finally, we conclude with a discussion of the status of our work, and future plans.

## 2. Keystone Science

As explicitly mentioned in the National Virtual Observatory White Paper (NVO 2001), a panchromatic census of active galactic nuclei is one of the original key science drivers for the creation of a NVO. Fundamentally, this particular science driver can be described as wanting to understand the formation and evolution of quasars and active galactic nuclei (AGN), and how they influence galaxy formation and evolution. Since it is rather broad in its scope, this particular science driver impacts many other sub-discipline and impacts many different wavelength regimes, making it an exceedingly interesting problem to tackle. Furthermore, since quasars are believed to be powered by super massive black holes and the data under study is inherently multi-wavelength, this work provides enormously interesting educational and public outreach opportunities.

### 2.1.   Science Drivers

The following list of questions summarizes the basic scientific drivers for the Skyserver project. Clearly, this list does not completely address all possible science projects that could utilize the resultant data and services, and, in fact, we feel this is one reason why our project is so important as it provides an enormously useful resource to the community. The rich dataset we will generate by federating the multi-wavelength surveys will be useful on its own for scientific exploration, as well as provide a valuable springboard for future targeted explorations using Chandra, XMM-Newton, GALEX, and SIRTF.

#### 2.1.1.  How do active galaxies form and evolve?

Although luminous AGN's are rather rare (~ 1% of all galaxies), they are extremely luminous, and serve as highly biased tracers of the underlying mass distribution, especially at higher redshifts (Osmer 1998). If as expected, their central engines are powered by black holes, the number density and spatial distribution place strong constraints on models of galaxy formation. A relatively recent development in this area has been photometric redshift estimation for quasi-stellar objects (Hatziminaoglou *et al.* 2000). Although still in its nascent stages, this new tool holds great promise in quantifying the statistical evolution of quasars and active galactic nuclei, as well as significantly improving the efficiency of future, targeted spectroscopic surveys.

### 2.1.2. What are the fundamental differences between the Radio-Loud and Radio-Quiet AGN populations?

Although Unification models have succeeded (at least so far) in presenting a single explanation for the observed differences between the radio-load and radio-quiet AGN populations, with an unbiased sample we will be able to place stronger constraints on any purely geometric differences. Currently, a bimodality in the radio to optical flux ratio is seen for optically selected samples, however, it is unknown whether this is an observational selection effect or indicative of something more fundamental (*e.g.,* see White *et al.* 2000 for a discussion of the properties of a radio selected quasar sample). A key point will be to look for this bimodality in a statistically significant radio selected sample, which might enable a detailed study of the evolution of radio-quiet quasars.

### 2.1.3. What is the relationship between Quasars and large-scale structure and how does it evolve with redshift?

Since Quasars are highly biased tracers of the underlying mass distribution, especially at high redshift, they can be used to locate possible overdensities in the galaxy distribution (Shanks and Boyle 1994). Quasars can be correlated with both optically and X-ray selected clusters in order to understand both the evolution of structure in the universe, as well as the effects of their environment on quasar evolution.

### 2.1.4. What is the dominant emission mechanism for Radio-Quiet Quasars?

Radio-Loud quasars appear to be dominated by non-thermal emission, however, radio-quiet quasars are more of a mystery, although thermal emission in the Optical and Ultraviolet is definitely more important than for the radio-loud sources (Padovani 1998). Any significant differences and their evolutionary trends can be used to place further limits on models for AGN unification.

### 2.1.5. Testing the AGN Unification Model?

Observational differences between the two types of active galactic nuclei are attributed to source-observer geometry in the majority of unification models. Therefore, Type 2 objects (narrow-lined) are fundamentally identical to Type 1 objects (broad-lined) and are only observed from a different position angle. These geometrical arguments, however, over predict the number of known Radio-Quiet Type 2 AGNs. A plausible explanation for this dearth of Type 2 Radio-Quiet AGNs is the original candidate selection technique; essentially Type 2 Radio-Quiet QSOs will be heavily obscured by the dusty torus, while radio selection will preferentially select radio-loud quasars (Padovani 1998).

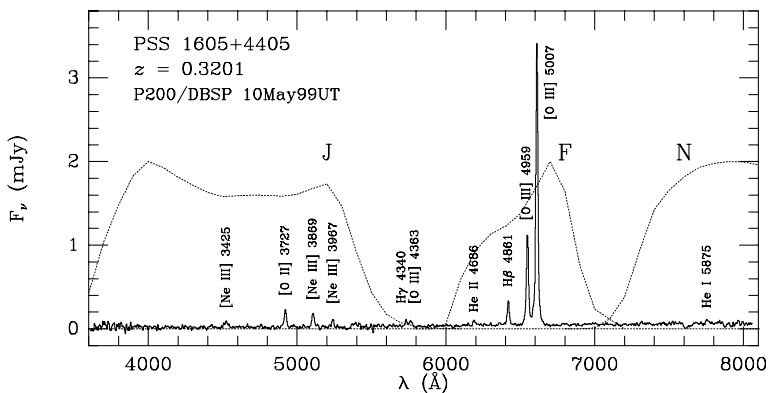

**Figure 1: A representative optical spectrum for one of the candidate Type 2 quasars. Note the strong, high-ionization lines. The DPOSS bandpasses are shown with dotted line: J (green), F (red), and N (near-IR). The [O III] emission lines dominate the flux in the red (F) band, leading to the color-outlier selection of these objects.**

Our multi-wavelength approach will circumvent these problems, as we will be able to select candidate Type 2 Radio-Quiet sources based on their multi-wavelength properties. As an example, in Figure 1, we show a Type 2 quasar candidate identified on the basis of its optical colors (which are odd, relative to normal galaxies, due to the strong emission lines). With spectrophotometric data we are currently obtaining, we will be able to optimally target specific areas within the multi-dimensional flux space our cross-identification catalog will occupy for candidate Type 2 quasars.

### 2.1.6. What is the relationship between Ultraluminous Infrared Galaxies and Quasars?

The absolute luminosity of Ultraluminous Infrared Galaxies (ULIRGs) or Hyperluminous Infrared Galaxies (HIGs), which are selected based on the ratio of their far-infrared to optical flux, are similar to QSOs (~ $10^{12}$ $L_*$), raising the possibility of a connection. We will be able to test any possible relationship between these two classes of sources from a more detailed multi-wavelength study along with follow-up spectroscopy. In addition, the federated catalog we will generate will be extremely useful to the general astronomical community in producing target lists for SIRTF follow-up spectroscopy in order to better understand these extremely luminous infrared sources.

### 2.1.7. What is the star formation history at low redshifts?

The evolution of the star formation rate (*i.e.,* the "Madau Plot") has become one of the most displayed figures in astronomy. The vast majority of the focus in trying to understand the star formation history of the universe, however, has been at the high redshift end. Equally important, and amazingly little understood is the low redshift star formation history, which due to the redshift-age relation is extremely important. We can provide a unique perspective on this question due to the extensive multi-wavelength nature of our data (UV–Far-IR morphology of local galaxies). For the brightest subset of the galaxies in our sample, we will be able to study the multi-wavelength variation in morphology, and understand how star formation and morphology are related. This project is facilitated by the morphological classifications that have been performed on the large ground based surveys (*e.g.,* Odewahn *et al.* 1998, Mazzarella *et al.* 1999).

## 2.2. Science Driver Overview

In attempting to answer the previous questions, we will construct the largest homogeneous quasar sample ever published. In addition we will produce a large, homogeneous catalog of BL Lac candidates, Type 2 Radio Quiet AGN candidates, and Ultraluminous Infrared Galaxy candidates. All of these candidates will invite follow-up ground and space based spectroscopy. Furthermore, with this new data, we can improve our understanding of the X-ray background and identify the likelihood of any new components. Finally, this cross-identification catalog can be used to improve our understanding of the local group by quantifying the galaxy distribution within the zone of avoidance in a truly multi-wavelength approach.

# 3. The Challenge: Handling the Data

In order to quantify the difficulty in federating astronomical survey data, consider the following. An all sky survey is 40,000 square degrees. If the typical pixel size for a particular survey of interest is 1 arcsecond, and the data is stored with two bytes (which provides a limited, but often sufficient dynamic range), approximately one Terabyte of data will be generated per band. While this quantity of data is not overly impressive nowadays (especially given that one can build a disk storage system to serve this data for under $10,000), we have multiple sky surveys at various wavelengths to federate, with significant overlap (for calibration, *etc.*). If a survey has higher spatial resolution, the data requirements increase significantly. Thus, our present problem (or good fortune, depending on your point of view) is one of handling hundreds of Terabytes. As we move into the time domain with synoptic surveys the challenge involves Petabytes of data.

A counterpoint to this calculation is that it was done for imaging data, not the catalog data, which is what most users want. While there is some validity to this argument, with the current calibration requirements (for astrometry and photometry), the data compression that results in moving from images to catalogs is not nearly as large as is generally assumed. Furthermore, in order to satisfy as many users as possible, a large number of different attributes are often measured, further increasing the size of the generated catalogs. Finally, many users will want to eventually touch the original imaging data, if for no other reason than to place upper limits on non-detections or to look for low surface brightness sources.

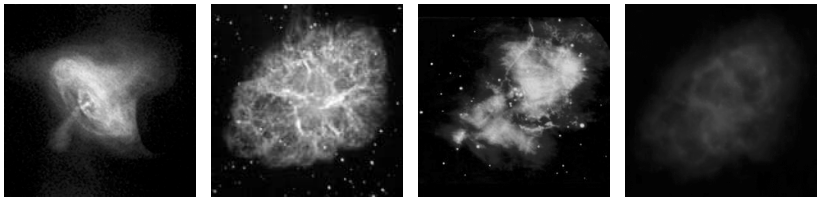

**Figure 2: X-ray, optical, infrared, and radio views of the nearby Crab Nebula, which is now in a state of chaotic expansion after a supernova explosion first sighted in 1054 A.D. by Chinese Astronomers. Although the images are not all on the same physical scale, a casual inspection clearly shows the significant morphological variations this one source displays in the different wavelength regimes. Figure Credit: NASA/CXC/SAO.**

Beyond just the physical quantity of data, federating astronomical data presents other, more subtle problems. First, astronomical data are highly distributed, not just nationally, but internationally. Second, astronomical data are very heterogeneous – metadata catalogs and standard transformations are vital to properly understand, transport, and dynamically process the data of interest. Third, astronomical data are NOT clean, and, therefore, traditional data-mining techniques cannot be readily applied. Fourth, the data of interest often spans multiple resolutions – typical factors of ten to a hundred in dynamic range are common. Thus unique associations are often not possible. Finally, astronomical data spans various physical phenomena – which are often uncorrelated, making the source association problem even more intractable (see, Figure 2 for a demonstration).

## 3.1. The Data

The data we are federating within the Skyserver project falls into three separate categories: ground-based data, primary space-based data, and supplemental space-based data.

### 3.1.1. Ground Based Data

The following ground based surveys form the foundation for our multi-wavelength cross-identification. The Digital Sky project, which is an NPACI funded project to serve as a technology demonstrator for a National Virtual Observatory, has already spent considerable effort exploring the problem of federating these datasets; thus we are not starting at ground zero.

The Digitized Palomar Observatory Sky Survey (DPOSS) is a multi-band optical survey ($g$, $r$, and $i$) of the entire northern sky generated by scanning the POSS-II photographic plate data. The final catalog is expected to contain approximately 50 million galaxies and 2 billion stars down to a limiting magnitude of $g < 22$. The survey is rapidly nearing completion for high galactic latitudes ($|b| > 15$).

The USNO-A2 catalog is a full-sky survey based on a re-reduction of the Precision Measuring Machine (PMM) scans of the original POSS-I and SERC photographic blue and red plates. It contains over five hundred million unresolved sources down to a limiting magnitude of B ~ $20^m$. Its inclusion in the catalog federation is primarily because it covers a larger area than the DPOSS survey, while it also has provided an early ability to test out algorithms and techniques while the final DPOSS calibrations are determined. Furthermore, the presence of multiple epochs will facilitate looking for variable sources.

The Two Micron All Sky Survey (2MASS) is a near-infrared ($J$, $H$, and $K_S$) all sky survey. The final catalog is expected to contain more than one million resolved galaxies, and more than three hundred million stars and other unresolved sources to a 10σ limiting magnitude of $K_S < 14.3$. The survey is currently completed, and the final, complete data release is pending a reprocessing of all imaging data.

The NRAO VLA Sky Survey (NVSS) is a publicly available, radio continuum survey covering the sky north of -40 degrees declination. The survey catalog contains over 1.8 millions discrete sources with total intensity and linear polarization image measurements (Stokes $I$, $Q$, and $U$) with a resolution of 45 arcseconds, and a completeness limit of about 2.5 mJy.

The Faint Images of the Radio Sky at Twenty-cm (FIRST) survey is a publicly available, radio snapshot survey that is scheduled to cover approximately ten thousand square degrees of the North and South Galactic Caps in 1.8 arcsecond pixels (currently, approximately eight thousand square degrees have been released). The survey catalog, when complete should contain around one million sources with a resolution of better than 1 arcseconds.

Currently, we are federating all publicly available data from the USNOA-2, 2MASS, and FIRST surveys. The next step is to add in the NVSS and DPOSS surveys. This last step is hampered, however, by the need for probabilistic associations (in the case of the NVSS data), and the continued reprocessing of the DPOSS data.

### 3.1.2. Primary Space Based Data

The following four space-based archives form the primary new data we are initially incorporating into the Skyserver project. Together with the core ground based data, these data extend the wavelength baseline of our sample to the far infrared and soft X-Ray. The extra wavelength coverage will allow for a complete sample of active galaxy candidates to be constructed for follow-up spectroscopy, based solely on their multi-wavelength properties.

The ROentgen SATellite (ROSAT) was an X-ray observatory that performed an all sky survey in the 0.1 to 2.4 keV range as well as numerous pointed observations. The full sky survey data has been publicly released, and includes two full sky catalogs. The first is the ROSAT All Sky Survey Bright Source Catalog (RASSBSC) which is a bright subsample (> 18,000) sources) of the all sky survey with positional accuracies > 6 arcseconds. The second catalog is the ROSAT All Sky Survey Faint Source Catalog (RASSFSC) which extends the Bright Source Catalog to fainter flux limits and includes a correspondingly larger number of sources (> 100,000). A supplemental catalog was compiled by White, Giommi, and Angelini (WGACAT; White *et al.* 1995) and includes all sources detected in the ROSAT pointed observations. Due to its serendipitous nature, it is extremely inhomogeneous, however it does provide ~62,000 additional sources.

The Infrared Space Observatory (ISO) was an astronomical satellite that operated at wavelengths from 2.5 to 240 microns. Both imaging and spectroscopic data are available over a large area of the sky.

The Infrared Astronomical Satellite (IRAS) performed an unbiased all sky survey at 12, 25, 60 and 100 microns, detecting ~350,000 high signal-to-noise infrared sources split between the faint and point source catalogs. A significantly larger number of sources (> 500,000) are included in the faint source reject file, which includes those sources that were below the flux threshold required for the faint source catalog.

The Einstein Observatory, originally the second High Energy Astrophysical Observatory (HEAO 2), was the first fully imaging X-ray telescope put into space, with an angular resolution of a few arcseconds and was sensitive over the energy range 0.2 to 3.5 keV.

### 3.1.3. Secondary Space Based Data

The following three surveys are supplemental to our primary scientific goals. They are included, however, since they are easily incorporated in the cross-identification catalog, and increase the scientific utility of the resultant data in two different ways. First, the ultraviolet imaging from UIT can be used to extend the wavelength baseline of our study of local galaxies and their star formation histories. Second, the ASCA and EUVE data can be used to preferentially identify interesting classes of objects (*i.e.,* Type-2 Quasars) for later follow-up based on their similarity to the multi-wavelength counterparts to these secondary space datasets. Finally, the UV data in these missions provides an invaluable guide as we prepare for the eventual inclusion of GALEX data in this analysis.

The Advanced Satellite for Cosmology (ASCA) is an X-Ray satellite that operated in the energy range 0.4 to 10 keV. The ASCA Solid-state Imaging Spectrometers (ASCA-SIS) catalog is a serendipitous search through publicly available archive data, which currently has around 1000 sources with positional accuracies of approximately 5 arcseconds. ASCA also performed several small area surveys. The spectral data from ASCA, however, is perhaps the most useful, and can be used to target objects from the ground-based catalog for follow-up spectroscopy that have similar properties as the ASCA sources.

The Extreme Ultraviolet Explorer (EUVE) conducted the first extreme ultraviolet (70 to 760 Angstroms) survey of the sky, followed by an ongoing program of pointed spectroscopy. Although the majority of EUVE sources are Galactic in nature, the unique wavelength coverage of this instrument necessitated its inclusion.

The Ultraviolet Imaging Telescope (UIT) flew on two separate shuttle missions, taking ultraviolet images of both stellar and extragalactic sources in two bands. The UV morphology provided by UIT on local galaxies will significantly enhance the multi-wavelength morphological studies of nearby galaxies.

# 4. The Skyserver Project

Broadly speaking, the Skyserver project has two primary goals. The first goal is to federate all, publicly available astronomical imaging datasets to facilitate knowledge discovery in the resulting rich dataset. The second goal is to scientifically verify the resulting data federation. Since the federation process is integral to the project's success, in this section, we thoroughly discuss the fundamental components of the process, followed by a higher-level discussion of the techniques that are available to evaluate the success of the source association process.

## *4.1. Federating the Data*

While the Skyserver project is certainly not the first multi-wavelength source cross-identification project, we feel the approach we have taken is vastly superior due to our large survey area (we will survey, at a minimum, several thousand square degrees, and most likely several times more), our fully probabilistic approach, and our open source development. In addition, all of our derived data products, as well as the original data we utilize, are non-proprietary.

The general problem we are tackling is, of course, complex for many reasons. First, it is a computationally challenging problem since there are billions of sources with varying spatial density across the sky that need to be associated. In addition, this process will need to be dynamic to allow different association algorithms to be employed. This last point is often forgotten, however, since the federation is multi-wavelength, a priori astrophysical knowledge is vitally important to the association process. Thus different user-defined, cross-identification algorithms will need to be supported. Finally, since astronomical sources are often variable (either spatially or temporally), the whole process becomes increasingly complex.

### 4.1.1. Hierarchical Triangular Mesh

The utilization of Hierarchical Triangular Mesh (HTM: Szalay & Brunner 1998) as a storage primitive has become increasingly common in large astronomical archives. One of the primary reasons for the widespread adoption of HTM was to simplify cross-archive data federation. As a result, we have built HTM support into our cross-identification algorithms in order to capitalize on the efficiencies that HTM presents. In the interest of completeness, we present a brief overview of this sky partitioning technique.

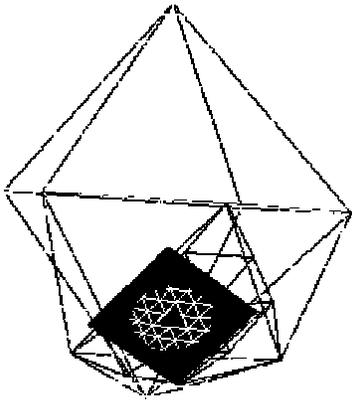

**Figure 3: An illustration of the HTM being utilized. In the figure, the black plane is the result of a proximity query, translated to Cartesian coordinates. All triangles to one side of the plane (which are easily visible) must be searched in more detail in order to completely determine which sources are within the requested distance from the original object.**

The HTM approach is based on a recursive division of the celestial sphere, starting with eight base regions, which are obtained by quartering the North and South hemispheres. The subdivision process involves creating four new areas that are roughly equal area by joining the midpoints of the three edges that bound the region. This process is recursively applied until a desired level of granularity is achieved (which can be allowed to vary spatially to account for variable densities). The hierarchical subdivisions can be very efficiently represented algorithmically by using quad-trees (*e.g.*, see Samet 1990).

Another simplification is achieved by the HTM naming scheme, where each triangle (or node on the quad-tree) is given a unique name. Since the tree is built hierarchically, the naming scheme is also completed hierarchically. The base regions are first named by the hemisphere in which they reside (N for North and S for South), and sequentially numbered in a predetermined order, thus the eight original regions are encoded as N0, N1, N2, N3, S0, S1, S2, and S3. The recursive subdivision process also lends itself naturally to continuing this naming scheme, where each of the four child regions is further numbered in a predetermined order. Thus, a given region N1 would have four children: N10, N11, N12, and N13, all of which are completely contained in the parent region N1.

This naming scheme enables queries to be made which quickly and easily grab the relevant portion of the sky, regardless of the coordinate system employed by the underlying archive. One final efficiency is gained by storing the source coordinates in Cartesian form, which is as a triplet: x, y, and z, where these values represent the position of the object on the unit sphere. While this results in slightly larger storage requirements per source, the resulting benefits are significant, as many of the computationally intensive queries in spherical space become considerably easier in Cartesian space (*e.g.*, see Figure 3). In addition, coordinate conversions, between different angular systems, for example, can be specified at run time allowing recalibrations to be seamlessly integrated.

### 4.1.2. Bayesian Probabilities of Association

Most wavelength cross-identifications are determined solely on spatial proximity. We are developing, however, a fully probabilistic approach that incorporates additional astrophysical a priori knowledge, such as multi-wavelength spectral classifications and redshift estimates, into a robust multi-wavelength source cross-identification. Our approach will allow for significantly fainter sources (often placed in reject files due to their low signal-to-noise characteristics) to be included in the overall analysis. This technique has its roots in the OPTID database (Lonsdale *et al.* 1998) that cross-matched the IRAS FSC with first generation optical photographic sky surveys. This technique simultaneously performs a maximum likelihood analysis in all available multi-wavelength data in order to determine the most likely source associations, as well as their probability of association. Additional prior knowledge such as the stellar distribution function, the number-magnitude relation, and multi-wavelength flux correlations can naturally be incorporated in the Bayesian approach.

The details of this technique are discussed in depth elsewhere (Lonsdale *et al.* 1998; Rutledge, Brunner, Prince, & Lonsdale 2000), however, in the interest of completeness, we present a brief overview of the basic premises. First, we calculate a likelihood of association between the objects in the two catalogs, which, for a given source (such as an X-Ray detection) is determined through a likelihood ratio ($LR_i$) that is assumed to be the product of a normalized probability distribution of a specific property of the target catalog objects (such as a specific flux measurement or flux ratios).

$$LR_i = \prod_j P_j(x_i)$$

where, for our purposes, the individual probabilities might be a function of both the relative spatial proximity of the sources in the different catalogs ($r_j$), as well as a specific flux ratios (*i.e.*, $f_x/f_o$). The individual source probabilities might be determined, for example, by

$$P_j = \frac{e^{\frac{-r_j^2}{2\sigma_j^2}}}{\sqrt{2\pi}\sigma_j} \frac{e^{\frac{-(f_t-f_j)^2}{2\sigma_t^2}}}{\sqrt{2\pi}\sigma_t}$$

where $f_j$ is the flux ratio of interest (*e.g.*, $f_x/f_o$) for the j$^{th}$ source, $f_t$ is a theoretical or model expectation for an a priori "best" value (*e.g.*, an average value for confirmed candidates), while $\sigma_j$ and $\sigma_t$ are determined from the observed spatial and flux ratio distributions respectively. In order to optimally target different classes of source counterparts, we merely need to tune the specific probability to preferentially select the candidates of interest (*e.g.*, Type-2 Quasars).

One final note about this technique, although important, the previous steps alone are not sufficient. The key step in this process is to compare the calculated likelihood ratio for the source field with suitably normalized background fields, which produces a reliability of identification:

$$R(LR) = \frac{N_{SourceField}(LR) - N_{RandomField}(LR)}{N_{SourceField}(LR)}$$

which is naturally a function of the determined likelihood ratio used. For each source in our catalog (or more likely, each counterpart which is within a specific spatial proximity of the target source), therefore, we can calculate the probability of identification

$$P_i = \frac{\frac{R_i}{1-R_i}\prod_j^N (1-R_j)}{S}$$

where $P_i$ is the probability of an identification for the i$^{th}$ source, and S is a normalization factor that guarantees the total probability calculated for a specific object (including the probability of no cross-catalog identification) will be unity.

$$S = \sum_i^N \frac{R_i}{1-R_i}\prod_j^N (1-R_j) + \prod_j^N (1-R_j)$$

where N is the number of possible counterparts (*i.e.*, the number of sources within the specified spatial proximity of the target source). In this manner, we can determine the specific probability of association for all sources in our catalog based on different astronomical insights. While not explicitly mentioned in this discussion, we can also use spectral classifications and photometric redshift indicators to both reduce the number of likely optical and near-infrared counterparts (*i.e.*, eliminate optical sources which are known to be x-ray quiet), as well as improve the actual probability reliability in addition to simple flux ratios (*i.e.*, though the utilization of statistical distance indicators).

### 4.1.3. Spectroscopic Classification and Photometric Redshifts

With the advent of the Hubble Deep Field (Williams *et al.* 1997), determining redshifts for objects from broadband photometry underwent a renaissance. Two complimentary approaches have gained popularity: spectral template fitting, and linear regression. The first technique (*e.g.,* Mazin and Brunner 2000, and references therein) uses model and observed spectral energy distributions to generate a redshift-magnitude grid. The observed magnitudes for a galaxy are then used to determine which spectral energy distribution from the grid of model values provides the best fit. The second approach (*e.g.,* Brunner 1997, Brunner *et al.* 1999) utilizes the inherent clustering of the galaxy distribution in multidimensional flux space (due to the discrete nature of allowed SED's) to empirically estimate redshifts. Both of these approaches have been shown to reliable estimate redshifts ($\delta_z \approx 0.06$) for Z < 1.0 (see, *e.g.*, Brunner *et al.* 1997) and ($\delta_z \approx 0.1$) for Z < 5 (Hogg *et al.* 1998).

For our purposes, however, these techniques have shortcomings. The traditional template fitting technique can only do as good as our knowledge of the spectral properties of the sources under consideration, also know as template incompleteness. On the other hand, the empirical technique requires a uniform set of training spectroscopic redshifts. Recently, however, a hybrid technique has been developed, in which templates are empirically derived from the dataset under analysis (Budavari *et al.* 2000). This approach is well suited to a federated optical, near-infrared dataset such as the one we are constructing (note that we will have at least 6 different flux measurements for the majority of our sources), since we have an overwhelmingly large number of galaxies with which to construct our eigen-spectral basis, which provides, simultaneously, both a redshift estimate and a spectral classification.

Basically, this technique utilizes the fact that galaxy spectra can be represented by a small number of spectral energy distributions (*i.e.,* the principal components or eigenspectra):

$$F = \sum_i a_i e_i$$

where $a_i$ is the i$^{th}$ eigenvalue, and $e_i$ is the i$^{th}$ eigenspectra (Connolly *et al.* 1999).

With a minor modification to the standard template photometric redshift estimation technique, we can now use the eigenbasis directly to estimate photometric redshifts. In particular, to determine a spectral classification for an astronomical source, we need to optimally select a spectral energy distribution from a grid of redshifted template spectra and constant stellar spectra. Algorithmically, we need to select the best fit to our data for different models, which can be easily done using chi-square fitting.

The appropriate eigenspectra templates can be derived either from existing spectral templates or model spectra, or they can be iteratively derived from the galaxy photometry directly. The second approach is what we are using in this project, as it can directly incorporate stars, galaxies, and quasars in a unified approach. Fundamentally, this technique treats the photometric observations as a very low-resolution spectrograph. However, since we have a very large number of sources at different redshifts, we actually have more than enough information to actually reconstruct the appropriate eigenspectral basis for our sample. As a starting point, we will utilize the Coleman, Weedman, and Wu galaxy spectral templates (Coleman *et al.* 1980), selected stellar templates from the Pickles library (Pickles *et al.* 1998), as well as synthetic quasar spectra (*cf.* Hatziminaoglou *et al.* 2000 for a more thorough discussion of generating quasar spectral energy distributions and quasar photometric redshifts).

Photometric redshift estimates can also utilize information from other wavelength regions, such as the radio to infrared flux ratio (Helou *et al.* 1985). However, the most important caveat to keep in mind when using photometric redshifts is that they should be used as statistical distance indicators. As a result, we define the probability density function, P(z), for an individual galaxy's redshift to be a Gaussian probability distribution function with mean (μ) given by the estimated photometric redshift and standard deviation (σ) defined by the estimated error in the photometric redshift (SubbaRao *et al.* 1996).

$$P(z) = \frac{1}{\sigma\sqrt{2\pi}} e^{\left(-\frac{(z-\mu)^2}{2\sigma^2}\right)}$$

The estimated error in the photometric redshift can be determined either by measuring the intrinsic dispersion in a given technique, or by generating bootstrap Monte-Carlo ensembles (*e.g.,* Brunner *et al.* 1999). In this way, we can reliable estimate the probability of a given galaxy being within a specific redshift range.

## *4.2.  Knowledge Discovery*

Now that the details of the data federation process have been explored, how do we anything useful with all of this data? This process is generally known as Knowledge Discovery in Databases (KDD). Fortunately for us, other groups are working on these problems. For example, the computational astrostastics group at Carnegie Mellon University has developed a cluster finding code that first identifies clusters in multidimensional data, and then assigns a value to all data vectors based on their distance from the clusters (Nichol 2001). This technique is extremely valuable in identifying and characterizing outliers in color space for subsequent follow-up (*e.g.,* Type-2 Quasars or High Redshift Quasars). Given an arbitrary parameter space, what we wish to determine is the clusters that naturally occur in this particular space. We also would like to find any isolated groups or holes in the clusters, as well as any isolated points or even isolated small clusters. This process of density estimation is one example of the power of collaborations between computer scientists, statisticians, and astronomers. We strongly encourage such partnerships as we feel they are vital to the successful implementation of virtual observatories. Other areas that should prove ripe for collaboration include performance improvements in existing algorithms, supervised and unsupervised classification, Bayesian Nets, genetic algorithms, and implementations that can utilize computational grids.

In Figure 4, the general topology of our planned framework is displayed, showing the connection between the data federation facilities and the astrostatistics services (such as the CMU density estimator). One promising mechanism for connecting the data facilities with the statistical algorithms is to employ web services. In this approach, the implemented algorithms are wrapped as SOAP servers, and are deployed within the framework.

Another potential model is to employ Enterprise JavaBeans to wrap the algorithms. This model could easily be deployed within a JavaSpace to take advantage of distributed compute services resulting in a dynamic computational grid. As a result, we are closely monitoring the JXTA project from SUN Microsystems, which has developed a peer-to-peer library in Java.

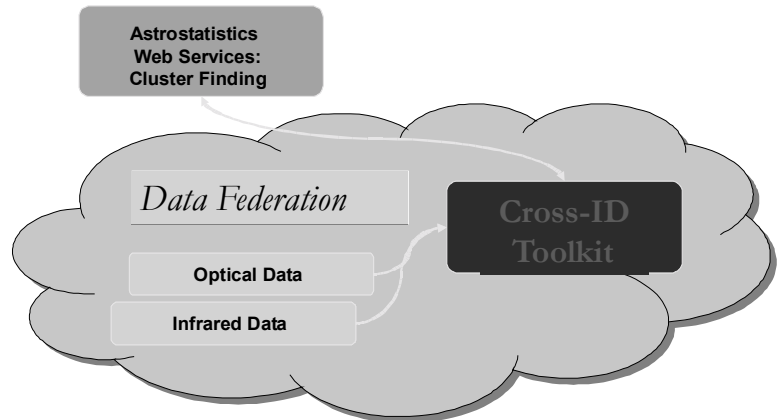

Figure 4: A diagram showing the topology of our planned framework for federating disparate data and exploring the resulting parameters spaces.

## 5.Conclusions

Currently the Skyserver facility consists of a single Microsoft SQL Server 2000 installation with the data of interested completely collocated. Clearly this is not an optimal solution, and goes against our stated design goals. As a result, one of our next tasks is to construct a mirror of the existing Skyserver installation, which will be located at Microsoft's Bay Area research campus under the direction of Jim Gray. Currently the data holdings include the SDSS EDR, the 2MASS second incremental release, the NVSS survey, the FIRST survey, the ROSAT survey, and the IRAS survey. In addition, all data federation is currently performed using spatial proximities (via HTM of course). Obviously this is not satisfactory, for the many reasons mentioned earlier in this proceedings, and probabilistic associations will be implemented, initially as stored procedures within SQL Server, but other options are also being explored.

While we currently have limited community access, primarily because we do not have the resources, we do plan on eventually opening up the entire facility, possibly within the context of the NASA Extragalactic Data Facility (NED).
This proceeding describes the Skyserver project, which is a Panchromatic Quasar Data Mining project. The science use cases for the Skyserver project form several key science drivers for the creation of Virtual Observatories. More information can be found at the project's webpage (www.skyserver.org). While important, this project is focusing on only one small part of the problem arising from the data Tsunami we are facing. Other projects are either on the drawing board, or underway. One example of a successful project that is working on the visualization aspects of a virtual observatory is virtualsky.org, which provides novel visualization of massive image archives. The combination of these different projects results in a whole that is greater than the sun of its parts, and provides enormous gains towards the ultimate construction of virtual observatories.

## ACKNOWLEDGEMENTS

We are grateful to all of our collaborators from around the world who share our vision, in particular the other members of the Digital Sky project. Particular thanks are given to George Djorgovski, Tom Prince, and Alex Szalay who have inspired and assisted with much of what is described in this proceeding, and to Jim Gray for both financial and moral support. In addition, we wish to thank NASA and NSF for their encouragement in difficult times, and both SUN Microsystems and Microsoft Research for their support. Finally, we would like to explicitly acknowledge financial support from NASA grants NAG5-10885 and NAG5-9482.